\documentclass[twocolumn]{revtex4}
\usepackage{amsthm, amssymb, graphicx, verbatim, tikz}

\usetikzlibrary{arrows,decorations.pathmorphing}

\usepackage{framed,color}

\def\t{^{\mbox{\tiny T}}}
\def\s{{\cal S}}
\def\g{{\cal G}}

\def\R{\mathbb{R}}
\def\Tr{{\rm Tr}}
\theoremstyle{plain} 
\newtheorem{theorem}{Theorem}
\newtheorem{lemma}{Lemma}
\newtheorem{requirement}{Postulate}
\definecolor{darkgreen}{rgb}{0,0.5,0}

\begin{document}

\title{Information-theoretic postulates for quantum theory
}
\author{Markus P.\ M\"uller}
\affiliation{Perimeter Institute for Theoretical Physics, 31 Caroline Street North, Waterloo, ON N2L 2Y5, Canada.}
\author{Llu\'\i s Masanes}
\affiliation{H.\ H.\ Wills Physics Laboratory, University of Bristol, Tyndall Avenue, Bristol, BS8 1TL, UK.}

\begin{abstract}
Why are the laws of physics formulated in terms of
complex Hilbert spaces? Are there natural and consistent modifications of quantum theory
that could be tested experimentally?
This book chapter gives a self-contained and accessible summary of our paper [New J.\ Phys.\ 13, 063001, 2011] addressing these questions, presenting the main
ideas, but dropping many technical details. We show that the formalism of quantum theory can be reconstructed from four natural postulates, which
do not refer
to the mathematical formalism, but only to the information-theoretic content of the physical theory.
Our starting point is to assume that there exist physical events (such as measurement outcomes) that happen probabilistically, yielding the
mathematical framework of ``convex state spaces''. Then, quantum theory can be reconstructed by assuming that (i) global states are determined
by correlations between local measurements, (ii) systems that carry the same amount of information have equivalent state spaces, (iii) reversible
time evolution can map every pure state to every other, and (iv) positivity of probabilities is the only restriction on the possible measurements.

\end{abstract}

\date{April 22, 2013}

\maketitle

\section{Introduction}
By all standards, quantum theory is one of the most successful theories of physics. It
provides the basis of particle physics, chemistry,
solid state physics, and it is of paramount importance for many technological achievements.
So far, all experiments have confirmed its universal validity in all parts of our physical world.
Unfortunately, quantum theory is also one of the most mysterious theories of physics.

In the text books, quantum theory is usually introduced by stating several abstract mathematical postulates:
\emph{States are unit vectors in a complex Hilbert space; probabilities are given by the Born rule; the Schr\"odinger
equation describes time evolution in closed systems}, to name just some of them. As many students recognize -- and experienced researchers
over the years sometimes tend to forget -- these postulates seem
arbitrary and do not have a clear meaning. It is true that they work very well and
are in accordance with experiments, but \emph{why are they true?} Why is nature described by these counterintuitive
laws of complex Hilbert spaces?

What at first sight seems to be a physically vacuous, philosophical question is in fact of high relevance to theoretical physics,
in particular for \emph{attempts to generalize quantum theory}.
There have been several attempts in the past to construct natural modifications of quantum theory -- either to set up experimental tests of quantum physics,
or to adapt it in a way which allows for easier unification with general relativity. However, modification of quantum theory
turned out to be a surprisingly difficult task.

A historical example is given by Weinberg's~\cite{Weinberg} non-linear modification of quantum theory.
Only a few months after his proposal was published, Gisin~\cite{Gisin} demonstrated that the resulting theory has an unexpected poisonous
property: it allows for superluminal signalling. It can be shown in general that other proposals of this kind must face a similar fate~\cite{NoSignaling}.
It seems as if the usual postulates of quantum theory
are intricately intertwined, in a way such that modification of one postulate makes the combination of the
others collapse into a physically meaningless -- or at least problematic -- theory.

One possible way to overcome this difficulty is to find alternative postulates for quantum theory that have a clear physical interpretation
and do not refer to the mathematical structure of complex Hilbert spaces. The search for simple operational axioms dates back to Birkhoff
and von Neumann~\cite{BvN}, and includes work by Mackey~\cite{gwmackey}, Ludwig~\cite{Ludwig}, Alfsen and Shultz~\cite{AlfsenShultzBook}
and many others. The advent of quantum information theory initiated new ideas and methods to approach this problem, resulting in the pioneering work
by Hardy~\cite{Hardy5}, and a recent wave of axiomatizations of quantum theory, including
Daki\'c and Brukner's work~\cite{Daki}, our result~\cite{MM}, the reconstruction by the Pavia group~\cite{GiulioAxioms},
alternative formulations by Hardy~\cite{Hardy11,Hardy12} and Zaopo~\cite{Zaopo}.

In this paper, we give a self-contained summary of our results in~\cite{MM}, where we derive the formalism of quantum theory
from four natural information-theoretic postulates. They can loosely be stated as follows:
\begin{enumerate}
\item The state of a composite system is characterized by the statistics of measurements on the individual components. 
\item All systems that effectively carry the same amount of information have equivalent state spaces.
\item Every pure state of a system can be transformed into every other by continuous reversible time evolution.
\item In systems that carry one bit of information, all measurements which give non-negative probabilities are allowed by the theory.
\end{enumerate}
Below, we show how to derive the usual formalism of quantum theory from these postulates. Surprisingly, the complex numbers
and Hilbert spaces pop out even though they are not mentioned in the postulates. This is true for all the axiomatization approaches
mentioned above, starting with Hardy's work~\cite{Hardy5}: these results allow us to gain a better understanding of the usual quantum formalism,
and resolve some of the mystery around ad hoc postulates like the Born rule.

Every axiomatization has its own benefits.  We think that the main advantage of our work~\cite{MM} -- as described in this paper -- is its
\emph{parsimony}: our postulates are close to a \emph{minimal} set of postulates for quantum theory. Accomplishing the goal of minimality
would mean to have a set of axioms such that dropping or weakening any one of the axioms will always yield new solutions in addition to quantum theory.
Currently, we do not know if we have actually achieved this goal, though we think that we are pretty close to it (this will be discussed
in more detail in Section~\ref{SecConclusions}). Our attempt to have as few assumptions as possible is also reflected in the background
assumptions: for example, we do not assume apriori that the composition of three systems into a joint system is associative,
or that pairs of generalized bits admit an analogue of a ``swap'' operation.

Our result suggests an obvious method to obtain natural modifications of quantum theory: \emph{drop or weaken one of the postulates,
and work out mathematically what the resulting set of theories looks like}. It is clear that minimality of the axioms (in the sense just described) is crucial for this method.
In contrast to the usual formulation of quantum theory, we know for sure that the corresponding alternative ``post-quantum'' theories
are consistent and do not allow for superluminal signalling as in Weinberg's approach. This is due to the fact that the no-signalling principle
is built in as a background assumption. In a way, those theories will be ``quantum theory's closest cousins'':
they are not formulated in terms of Hilbert spaces, but share as many characteristic features with quantum theory as possible.

As the simplest possible modification, suppose we drop the word ``continuous'' from Postulate 3 -- that is, we allow for discrete
reversible time evolution. Then another solution in addition to quantum theory appears: in this theory,
states are probability distributions, and reversible time evolution is given by permutations of outcomes. This is exactly
\emph{classical probability theory} on discrete sample spaces. It turns out to be the unique additional solution in this case.

\section{What do we mean by ``quantum theory''?}
When talking about axiomatizing quantum theory, there is sometimes confusion about what we actually mean by it.
The term ``quantum theory'' arouses association with many different aspects of physics that are usually treated in
quantum mechanics text books, such as particles, the hydrogen atom, three-dimensional position and momentum space
and many more.

However, a more careful 
definition should apply here. As an analogy, consider the theory of statistical mechanics. This theory
consists of an application of probability theory to mechanics, which means in particular that abstract probability theory can be studied
detached from statistical physics -- and this has been done in mathematics for a long time.

Similarly, we can consider quantum mechanics to be a combination of an abstract probabilistic theory -- \emph{quantum theory} --
and classical mechanics. Abstract quantum theory can be studied detached from its mechanical realization;
the main difference to the previous example lies in the historical fact that the development of quantum mechanics preceded
that of abstract quantum theory.
In this terminology, we understand by ``quantum theory'' the statement
that
\begin{itemize}
\item states are vectors (resp.\ density matrices) in a complex Hilbert space,
\item probabilities are computed by the Born rule resp.\ trace rule,
\item the possible reversible transformations are the unitaries,
\item measurements are described by projection operators, and thus observables are
given by self-adjoint matrices.
\end{itemize}
The ``classical mechanics'' part, on the other hand, determines the type of Hilbert space to consider (such as $L^2(\R^3)$), the
choice of ``Hamiltonians'' $H$ which generate the time evolution, $U(t)=\exp(i H t)$, and
the choice of initial states of that time evolution.
This conceptual distinction has proven particularly useful
in the development of quantum information theory. It seems that this distinction was
always implicit when expressing the desire to ``quantize'' any classical physical theory,
that is, to combine it with abstract quantum theory.

Thus, since we are aiming for a reconstruction of abstract quantum theory, we will not refer to position, momentum, or
Hamiltonians in this paper. Instead, we only use the notions of abstract probability theory: of events, happening with certain
probabilities, and of transformations modifying the probabilities. Furthermore, we restrict our analysis to finite-dimensional systems:
we argue that the main mystery is \emph{why to have a complex Hilbert space at all}. If this is understood in finite dimensions, it
seems only a small conceptual (though possibly mathematically challenging) step to guess the correct infinite-dimensional generalizations.

Since we presuppose probabilities as given, we also do not address the question where these probabilities come from.
Hence we also ignore the question about what happens in a quantum measurement, and all other interpretational mysteries
encompassing the formulation of quantum theory. Instead, we restrict ourselves to ask how the mathematical formalism
of quantum theory can be derived from simpler postulates, and what possible modifications of it we might hope to find in nature.

\begin{framed}
\noindent
\textbf{Questions that we would like to address:}
\begin{itemize}
\item How can we understand (that is, derive) the complex Hilbert space formalism from
simple operational assumptions on probabilities?
\item What other probabilistic theories are operationally closest to quantum theory?
\end{itemize}
\textbf{Questions/problems that we do \emph{not} address:}
\begin{itemize}
\item How should we interpret  ``probability'', and where does it come from?
\item The measurement problem.
\item Interpretation of quantum mechanics.
\end{itemize}
\end{framed}

In order to formulate our postulates, we work with a simple and general framework encompassing
all conceivable ways to formulate physical theories of probability: this is the framework of \emph{generalized
probabilistic theories}.

\section{Generalized probabilistic theories}
\label{gpt}
Classical probability theory (abbreviated CPT henceforth) is used to describe processes which are not
deterministic.
This is achieved by assuming a particular mathematical structure:
a probability space with a unique fixed probability measure, which is used to assign probabilities to all random variables.
The framework of generalized probabilistic theories~\cite{tomo,Hardy5,Barn,Barr,gwmackey,BvN,darian} generalizes this approach in a simple way.
We will now give a brief introduction to this framework, built on general considerations of what constitutes an experiment in physics.
For more detailed introductions, we refer the reader to~\cite{Barr,Barn}, and for nice presentations of the main ideas to~\cite{br, BruknerRules}.

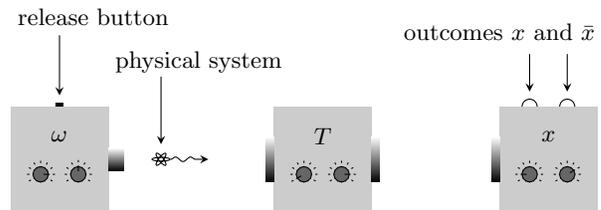
\begin{figure}
\begin{tikzpicture}[>=stealth]


\fill[fill=white!80!black] (.5,0) rectangle +(1.3,1.4); \shade[top
color=white, bottom color=black] (1.808,.55) rectangle +(.2,.3);
\draw (1.15,1) node {$\omega$};

\fill[fill=black] (1.1,1.4) rectangle +(.1,.05); \draw[<-]
(1.15,1.55) --+(0,.8); \draw (1.6,2.6) node {release button};

\draw [xshift=2.5cm, yshift=.7cm] (0,0) ellipse (.12cm and .025cm)
(0,0) [rotate=52] ellipse (.10cm and .03cm); \draw [xshift=2.5cm,
yshift=.7cm] (0,0) [rotate=128] ellipse (.10cm and .03cm); \draw
[->,decorate,
decoration={snake,amplitude=.3mm,segment length=2mm,post length=.8mm}] (2.63,.7) --+(.5,0);

\draw[<-] (2.5,.9) --+(0,.9); \draw (3,2) node {physical system};

\filldraw[fill=white!40!black, draw=black]  (.9,.5) circle (.1cm);
\draw (.9,.5)++(0:.04cm) --++(0:.08cm) ++(0:.04cm) --++(0:.03cm);
\draw (.9,.5)++(30:.1cm) --++(30:.01cm) ++(30:.04cm) --++(30:.03cm);
\draw (.9,.5)++(60:.1cm) --++(60:.01cm) ++(60:.04cm) --++(60:.03cm);
\draw (.9,.5)++(90:.1cm) --++(90:.01cm) ++(90:.04cm) --++(90:.03cm);
\draw (.9,.5)++(120:.1cm) --++(120:.01cm) ++(120:.04cm) --++(120:.03cm);
\draw (.9,.5)++(150:.1cm) --++(150:.01cm) ++(150:.04cm) --++(150:.03cm);
\draw (.9,.5)++(180:.1cm) --++(180:.01cm) ++(180:.04cm) --++(180:.03cm);
\draw (.9,.5)++(210:.1cm) --++(210:.01cm) ++(210:.04cm) --++(210:.03cm);
\draw (.9,.5)++(240:.1cm) --++(240:.01cm) ;
\draw (.9,.5)++(270:.1cm) --++(270:.01cm) ;
\draw (.9,.5)++(300:.1cm) --++(300:.01cm) ;
\draw (.9,.5)++(330:.1cm) --++(330:.01cm) ++(330:.04cm) --++(330:.03cm);

\filldraw[fill=white!40!black, draw=black]  (1.4,.5) circle (.1cm);
\draw (1.4,.5)++(0:.1cm) --++(0:.01cm) ++(0:.04cm) --++(0:.03cm);
\draw (1.4,.5)++(30:.1cm) --++(30:.01cm) ++(30:.04cm) --++(30:.03cm);
\draw (1.4,.5)++(60:.1cm) --++(60:.01cm) ++(60:.04cm) --++(60:.03cm);
\draw (1.4,.5)++(90:.04cm) --++(90:.08cm) ++(90:.04cm) --++(90:.03cm);
\draw (1.4,.5)++(120:.1cm) --++(120:.01cm) ++(120:.04cm) --++(120:.03cm);
\draw (1.4,.5)++(150:.1cm) --++(150:.01cm) ++(150:.04cm) --++(150:.03cm);
\draw (1.4,.5)++(180:.1cm) --++(180:.01cm) ++(180:.04cm) --++(180:.03cm);
\draw (1.4,.5)++(210:.1cm) --++(210:.01cm) ++(210:.04cm) --++(210:.03cm);
\draw (1.4,.5)++(240:.1cm) --++(240:.01cm) ;
\draw (1.4,.5)++(270:.1cm) --++(270:.01cm) ;
\draw (1.4,.5)++(300:.1cm) --++(300:.01cm) ;
\draw (1.4,.5)++(330:.1cm) --++(330:.01cm) ++(330:.04cm) --++(330:.03cm);


\fill[fill=white!80!black] (4,0) rectangle +(1.3,1.4);

\shade[top color=white, bottom color=black] (5.3,.4) rectangle
+(.1,.6); \shade[top color=white, bottom color=black] (4,.4)
rectangle +(-.1,.6);

\draw (4.65,1) node {$T$};
\filldraw[fill=white!40!black,draw=black]  (4.4,.5) circle (.1cm);
\draw (4.4,.5)++(0:.1cm) --++(0:.01cm) ++(0:.04cm) --++(0:.03cm);
\draw (4.4,.5)++(30:.1cm) --++(30:.01cm) ++(30:.04cm)
--++(30:.03cm); \draw (4.4,.5)++(60:.1cm) --++(60:.01cm)
++(60:.04cm) --++(60:.03cm); \draw (4.4,.5)++(90:.1cm)
--++(90:.01cm) ++(90:.04cm) --++(90:.03cm); \draw
(4.4,.5)++(120:.1cm) --++(120:.01cm) ++(120:.04cm)
--++(120:.03cm); \draw (4.4,.5)++(150:.1cm) --++(150:.01cm)
++(150:.04cm) --++(150:.03cm); \draw (4.4,.5)++(180:.1cm)
--++(180:.01cm) ++(180:.04cm) --++(180:.03cm); \draw
(4.4,.5)++(210:.04cm) --++(210:.08cm) ++(210:.04cm)
--++(210:.03cm); \draw (4.4,.5)++(240:.1cm) --++(240:.01cm) ;
\draw (4.4,.5)++(270:.1cm) --++(270:.01cm) ; \draw
(4.4,.5)++(300:.1cm) --++(300:.01cm) ; \draw (4.4,.5)++(330:.1cm)
--++(330:.01cm) ++(330:.04cm) --++(330:.03cm);

\filldraw[fill=white!40!black, draw=black]  (4.9,.5) circle
(.1cm); \draw (4.9,.5)++(0:.04cm) --++(0:.08cm) ++(0:.04cm)
--++(0:.03cm); \draw (4.9,.5)++(30:.1cm) --++(30:.01cm)
++(30:.04cm) --++(30:.03cm); \draw (4.9,.5)++(60:.1cm)
--++(60:.01cm) ++(60:.04cm) --++(60:.03cm); \draw
(4.9,.5)++(90:.1cm) --++(90:.01cm) ++(90:.04cm) --++(90:.03cm);
\draw (4.9,.5)++(120:.1cm) --++(120:.01cm) ++(120:.04cm)
--++(120:.03cm); \draw (4.9,.5)++(150:.1cm) --++(150:.01cm)
++(150:.04cm) --++(150:.03cm); \draw (4.9,.5)++(180:.1cm)
--++(180:.01cm) ++(180:.04cm) --++(180:.03cm); \draw
(4.9,.5)++(210:.1cm) --++(210:.01cm) ++(210:.04cm)
--++(210:.03cm); \draw (4.9,.5)++(240:.1cm) --++(240:.01cm) ;
\draw (4.9,.5)++(270:.1cm) --++(270:.01cm) ; \draw
(4.9,.5)++(300:.1cm) --++(300:.01cm) ; \draw (4.9,.5)++(330:.1cm)
--++(330:.01cm) ++(330:.04cm) --++(330:.03cm);


\fill[fill=white!80!black] (7,0) rectangle +(1.3,1.4); \shade[top
color=white, bottom color=black] (7,.4) rectangle +(-.1,.6);

\draw[<-] (7.4,1.6) --+(0,.5); \draw[<-] (7.9,1.6) --+(0,.5);
\draw (7,2.4) node {outcomes $x$ and $\bar{x}$};

\draw (7.5,1.4) arc (0:180:.1cm); \draw (8,1.4) arc (0:180:.1cm);

\draw (7.65,1) node {$x$};
\filldraw[fill=white!40!black,draw=black]  (7.4,.5) circle (.1cm);

\draw (7.4,.5)++(0:.1cm) --++(0:.01cm) ++(0:.04cm) --++(0:.03cm);

\draw (7.4,.5)++(30:.1cm) --++(30:.01cm) ++(30:.04cm)
--++(30:.03cm);

\draw (7.4,.5)++(60:.1cm) --++(60:.01cm) ++(60:.04cm)
--++(60:.03cm);

\draw (7.4,.5)++(90:.1cm) --++(90:.01cm) ++(90:.04cm)
--++(90:.03cm);

\draw (7.4,.5)++(120:.1cm) --++(120:.01cm) ++(120:.04cm)
--++(120:.03cm);

\draw (7.4,.5)++(150:.1cm) --++(150:.01cm) ++(150:.04cm)
--++(150:.03cm);

\draw (7.4,.5)++(180:.04cm) --++(180:.08cm) ++(180:.04cm)
--++(180:.03cm);

\draw (7.4,.5)++(210:.1cm) --++(210:.01cm) ++(210:.04cm)
--++(210:.03cm);

\draw (7.4,.5)++(240:.1cm) --++(240:.01cm);

\draw (7.4,.5)++(270:.1cm) --++(270:.01cm);

\draw (7.4,.5)++(300:.1cm) --++(300:.01cm);

\draw (7.4,.5)++(330:.1cm) --++(330:.01cm) ++(330:.04cm)
--++(330:.03cm);

\filldraw[fill=white!40!black, draw=black]  (7.9,.5) circle
(.1cm);

\draw (7.9,.5)++(0:.1cm) --++(0:.01cm) ++(0:.04cm) --++(0:.03cm);

\draw (7.9,.5)++(30:.04cm) --++(30:.08cm) ++(30:.04cm)
--++(30:.03cm);

\draw (7.9,.5)++(60:.1cm) --++(60:.01cm) ++(60:.04cm)
--++(60:.03cm);

\draw (7.9,.5)++(90:.1cm) --++(90:.01cm) ++(90:.04cm)
--++(90:.03cm);

\draw (7.9,.5)++(120:.1cm) --++(120:.01cm) ++(120:.04cm)
--++(120:.03cm); \draw (7.9,.5)++(150:.1cm) --++(150:.01cm)
++(150:.04cm) --++(150:.03cm); \draw (7.9,.5)++(180:.1cm)
--++(180:.01cm) ++(180:.04cm) --++(180:.03cm); \draw
(7.9,.5)++(210:.1cm) --++(210:.01cm) ++(210:.04cm)
--++(210:.03cm); \draw (7.9,.5)++(240:.1cm) --++(240:.01cm) ;
\draw (7.9,.5)++(270:.1cm) --++(270:.01cm) ; \draw
(7.9,.5)++(300:.1cm) --++(300:.01cm) ; \draw (7.9,.5)++(330:.1cm)
--++(330:.01cm) ++(330:.04cm) --++(330:.03cm);

\end{tikzpicture}
\caption{{\bf General experimental set up.} From left to right there are the preparation, transformation and measurement devices. As
soon as the release button is pressed, the preparation device outputs a physical system in the state specified by the knobs. The next device performs the transformation specified by its knobs (which in particular can be ``do nothing"). The device on the right performs the measurement specified by its knobs, and the outcome \mbox{($x$ or $\bar{x}$)} is indicated by the corresponding light.
\label{f11}}
\end{figure}

In order to set up a common picture, we consider Figure~\ref{f11} as the model for what constitutes a physical experiment. This is just an illustration: the events that we describe
may as well be natural processes that happen without human or technological intervention.

The main idea (cf.~Figure~\ref{f11}) is that physical systems can cause objective events which we call ``measurement outcomes'' -- for example
clicks of detectors. We say that two systems are in the same state $\omega$ if all outcome
probabilities of all possible measurements are the same. In order to test this empirically, we always assume that
we can prepare a physical system in a given state as often as we want. That is, we may think of a \emph{preparation
device} which produces a physical system in a particular state.

\subsection{States and measurements}
\label{SubsecStatesMeasurements}
Single outcomes of measurements are called \emph{effects}, and are denoted by uppercase letters such as $E$.
The probability of obtaining outcome $E$, if measured on state $\omega$, will be denoted $E(\omega)$. This way,
effects become maps from states to probabilities in $[0,1]$.

What can we say about the set of all possible states $\omega$ in which a given system can be prepared? Suppose we
have two preparation devices; one of them prepares the system in some state $\omega$, the other one prepares it
in some state $\varphi$. Then we can use these devices to construct a new device, which tosses a coin,
and then prepares either state $\omega$ with probability $p\in[0,1]$, or state $\varphi$ with probability $1-p$.
We denote this new state by
\[
   \omega':=p\omega+(1-p)\varphi.
\]
Clearly, if we apply a measurement on $\omega'$, we get outcome $E$ with probability
\[
   E(\omega')=p E(\omega)+(1-p) E(\varphi).
\]
Thus, by this construction, we see that states $\omega$ become elements of an affine space, and effects $E$ are affine maps.
The set of all possible states -- called the \emph{state space $\s$} -- will be a subset of this affine space. We have just
seen that $\omega\in\s$ and $\varphi\in\s$ imply $p\omega+(1-p)\varphi\in\s$ if $0\leq p \leq 1$; that is,
state spaces are convex sets (similar reasoning is given in~\cite{Holevo,Hardy5,Barr}).

In principle, state spaces can be infinite-dimensional (and in fact, in many physical situations, they are). However, in this
paper, we will only consider finite-dimensional state spaces. Then, states $\omega$ are determined by finitely many coordinates,
and we may use this to construct a more concrete representation of states. Denote the dimension of a state space $\s$ by $d$.
Then, by choosing $d$ affinely independent effects $E_1,\ldots, E_d$, the probabilities $E_1(\omega),\ldots, E_d(\omega)$ determine
$\omega$ uniquely. We now use the representation
\begin{equation}
   \omega=\left[\begin{array}{c} 1 \\ E_1(\omega) \\ E_2(\omega) \\ \vdots \\ E_d(\omega)\end{array}\right] =: 
   \left[\begin{array}{c} 1 \\ \omega_1 \\ \omega_2 \\ \vdots \\ \omega_d\end{array}\right] \in \s\subset \R^{d+1}.
   \label{eqRepStates}
\end{equation}
The choice of $E_1,\ldots,E_d$ is arbitrary, subject only to the restriction that they are affinely independent. We call a set of effects
with this property \emph{fiducial}, and we refer to $E_1(\omega),\ldots, E_d(\omega)$ as \emph{fiducial outcome probabilities}~\cite{Hardy5}.
The component $\omega_0:=1$ has been introduced  for calculational convenience: it allows us to write the affine effects $E$
as \emph{linear} functionals on the larger space $\R^{d+1}$. It will also turn out to be particularly useful in calculations involving
composite state spaces.

In the following, we will assume that state spaces $\s$ are topologically closed and bounded, i.e.\ compact (for
a physical motivation see~\cite{MM}). The extremal points of the convex set $\s$ will be called \emph{pure states};
these are states $\omega$ which cannot be written as mixtures $p \varphi+(1-p)\varphi'$ of other states $\varphi\neq \varphi'$
with $0<p<1$. It follows from the compactness of $\s$ that every state can be written as a convex combination of at most $d+1$
pure states~\cite{convex_book}.

Measurements with $n$ outcomes are described by a collection of $n$ effects $E_1,E_2,\ldots, E_n$ with the property
$E_1(\omega)+E_2(\omega)+\ldots+E_n(\omega)=1$ for all states $\omega$. This expresses the fact that outcome $i$
happens with probability $E_i(\omega)$, and the total probability is one. Note that two effects $E$ and $F$ can only be
part of the same measurement if $E(\omega)+F(\omega)\leq 1$ for all states $\omega$. Sets of fiducial effects (as introduced
above) do not necessarily have this property. A single effect $E$ is always part of a measurement with two outcomes $E$ and
$\bar E$, where $\bar E(\omega):=1-E(\omega)$.

\begin{figure}[!hbt]
\begin{center}
\includegraphics[angle=0, width=8.5cm]{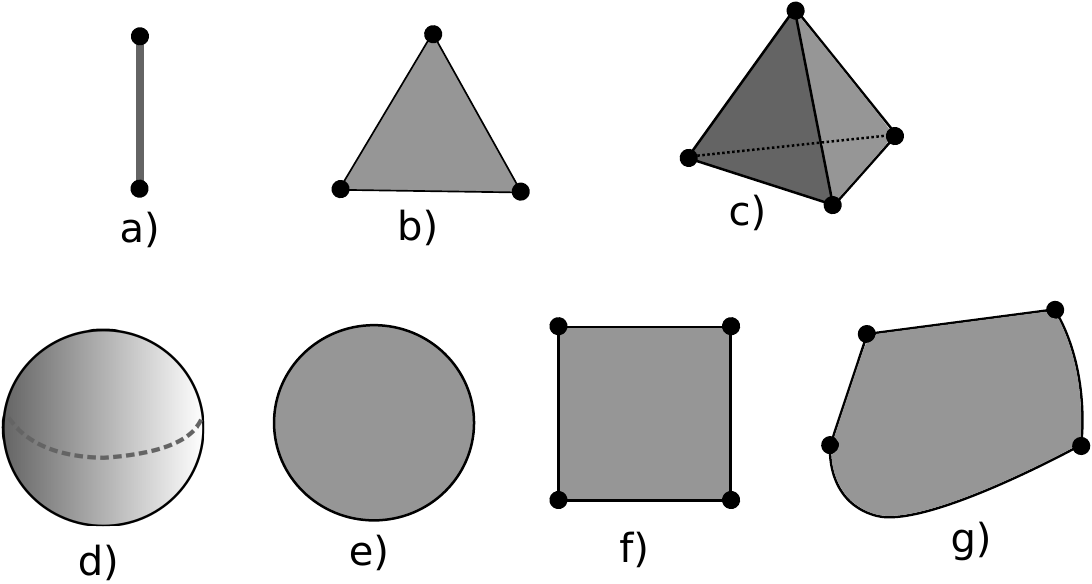}
\caption{Examples of convex state spaces: a) is a classical bit, b) and c) are classical $3$- and $4$-level systems, d)  is a quantum bit, e) is the projection of a qubit, f) and g) are
neither classical nor quantum.
Note that quantum $n$-level systems for $n\geq 3$ are \emph{not} balls.}
\label{fig_convexsets}
\end{center}
\end{figure}

Figure~\ref{fig_convexsets} gives some examples of convex state spaces. First, consider a classical bit, which is described within
CPT. We can think of a coin which shows either heads or tails; in general, it can be in one of those
configurations with some probability. The probability $p$ of showing heads determines the state uniquely, since the tails probability
must be $1-p$. Thus, $p\in [0,1]$ is a fiducial probability; recalling~(\ref{eqRepStates}), we can represent states as $\omega=[1,p]\t$.
This yields a one-dimensional state space, with two pure states $[1,0]\t$ and $[1,1]\t$, corresponding to coins which deterministically
show heads or tails. It is depicted in Figure~\ref{fig_convexsets}a).

Similarly, classical $n$-level systems have states which correspond to probability distributions $p_1,\ldots, p_n$. Since $p_n=1-(p_1+\ldots+p_{n-1})$,
the numbers $p_1,\ldots,p_{n-1}$ are fiducial outcome probabilities, yielding states $\omega=[1,p_1,\ldots,p_{n-1}]\t$.
Geometrically, the resulting state spaces are simplices. They are depicted
in Figure~\ref{fig_convexsets}b) and c) for $n=2$ and $n=3$.

Quantum systems look very different: as it is well-known, states of quantum $2$-level systems, i.e.\ qubits, can be parametrized by a vector $\vec r\in\mathbb{R}^3$
with $|\vec r|\leq 1$,
such that every density matrix can be written $\rho=(\mathbf{1}+\vec r \cdot\vec \sigma)/2$, with $\vec \sigma=(\sigma_x,\sigma_y,\sigma_z)$ the Pauli
matrices. Thus, we can use the vector $[1,r'_x,r'_y,r'_z]\t$ to represent states, where $r'_i:=(1+r_i)/2$ is the probability to measure ``spin up'' in $i$-direction.
This state space is the famous (slightly reparametrized) Bloch ball, cf. Figure~\ref{fig_convexsets}d).

Figure~\ref{fig_convexsets}e) shows a state space which is a projection of the Bloch ball: it corresponds to the effective state space that we
obtain if, for some reason, spin measurements in $z$-direction are physically impossible to implement, with states $\omega=[1,r'_x,r'_y]\t$.
The square state space in Figure~\ref{fig_convexsets}f)
describes a system for which there exist two independent effects, say $X$ and $Y$, that can yield probabilities $X(\omega)$ and $Y(\omega)$ in $[0,1]$
arbitrarily and independently from each other. States will be of the form $\omega=[1,\omega_x,\omega_y]\t$, with $\omega_x=X(\omega)$ and
$\omega_y=Y(\omega)$.

Consider the two yes-no-measurements which correspond to the effects $X$ and $Y$; we can interpret these as spin measurements
in two orthogonal directions, with ``yes''-outcome $X$ or $Y$ for ``spin up'', and ``no''-outcome $\bar X$ or $\bar Y$ for ``spin down''. 
If we perform either one of these measurements on the state $\omega=(1,1,1)$, then we will
get the ``yes''-outcome with unit probability -- and this is true for both measurements. If we consider the analogous measurements on the circle state space,
we see that the corresponding behavior becomes impossible: if one of the spin measurements yields outcome ``yes'' with certainty, then
the other spin measurement must give outcome ``yes'' with probability $1/2$. This follows from $r_x^2+r_y^2\leq 1$.

Thus, the circle state space shows a form of \emph{complementarity}, which is not present in the square state space. As this example illustrates,
the state space of a physical system determines many of its information-theoretic properties. Given a description of the state space
$\s$, we can also determine the set of all linear functionals which map states to the unit interval $[0,1]$, that is, the candidates for
possible effects. However, not all of them may be possible to implement in physics: maybe some of them are ``forbidden'', similarly
as superselection rules forbid some superpositions in quantum mechanics.
Therefore, to every given state space $\s_A$, there is
a set of ``allowed effects'' which are interpreted as those that can actually be physically performed.

We introduce some notions which will be useful later:
A set of states $\omega_1,\ldots,\omega_n$ is called \emph{distinguishable}
if there is a measurement with outcomes represented by effects $E_1,\ldots,E_n$, such that
$E_i(\omega_j)=\delta_{ij}$, which is $1$ if $i=j$ and $0$ otherwise.
The interpretation is that we can build a device which perfectly distinguishes the different states $\omega_j$. Given a physical system $A$, we define
the \emph{capacity} $N_A$ as the maximal size of any set of distinguishable states $\omega_1,\ldots,\omega_n\in \s_A$. A measurement
which is able to distinguish $N_A$ states (that is, as much as possible) will be called \emph{complete}.
For a quantum state space, $N_A$ equals the dimension of the underlying complex Hilbert space.

We denote the real vector space which carries $\s_A$ by $V_A$. Then effects are elements of the dual space $V_A^*$.
For a quantum $N$-level system, $V_A$ is the real vector space of Hermitian $N\times N$-matrices with complex entries.
Following Wootters and Hardy~\cite{Wootters, Hardy5}, we also use the notation $K_A:=\dim V_A=\dim(\s_A)+1$, that is the number
of degrees of freedom that is necessary to describe an unnormalized state.
For a qubit, for example, we have $N_A=2$, but $K_A=4$. In quantum theory,
$K_A=N_A^2$ equals the number of independent real parameters in a density matrix (dropping normalization). In classical probability
theory, we always have $K_A=N_A$.

\subsection{Transformations}
\label{s.trans}
A transformation is a map $T$ which takes a state to another state. Which transformations are actually possible is a question of physics.
However, there are certain minimal assumptions that every transformation must necessarily satisfy in order to be physically meaningful
in the context of convex state spaces. First, transformations must respect
probabilistic mixtures -- that is,
\[
   T(p\omega+(1-p)\varphi)=p T(\omega)+(1-p)T(\varphi).
\]
This is because both sides of the equation can be interpreted as the result of randomly preparing $\omega$ or $\varphi$ (with probabilities $p$ resp.\ $1-p$)
and applying the transformation $T$. Thus, transformations (from one system to itself) are affine maps which map a state space
$\s_A$ into itself; we can always assume that they are linear maps $T:V_A\to V_A$.

If both $T$ and $T^{-1}$ are physically allowed transformations, we call $T$ \emph{reversible}. The set of reversible transformations on a
physical system $A$ is a group $\g_A$. For physical reasons, we assume that $\g_A$ is topologically closed, hence a compact group~\cite{matrix_groups} (it may be
a finite group).

Reversible transformations map a state space bijectively onto itself -- hence they are symmetries of the state space. For example,
in quantum theory, reversible transformations are the unitary conjugations, $\rho\mapsto U \rho U^\dagger$. In the Bloch ball representation
of the qubit (as in Figure~\ref{fig_convexsets}d)), these maps are represented as rotations, such that the group of reversible transformations
is isomorphic to $SO(3)$.

However, as this example also shows, not all symmetries are automatically allowed reversible transformations: a reflection in the Bloch ball is a symmetry,
but it is not an allowed transformation (in the density matrix picture, it would correspond to an anti-unitary map).

In summary, for what follows, a physical system $A$ is specified by three mathematical objects: the state space $\s_A$,
the group of reversible transformations $\g_A$ (which is a compact subgroup of all symmetries of $\s_A$), and a set
of physically allowed effects. The latter will not be given a particular notation, but we assume that the set of allowed effects
is topologically closed. For obvious physical reasons, if $E$ is an allowed effect and $T\in\g_A$, then $E\circ T$ is an allowed
effect; similarly, convex combinations of allowed effects are allowed.

\subsection{Composite systems}\label{multip_ns}
If we are given two physical systems $A$ and $B$, we would like to define a \emph{composite system} $AB$ which
is also a physical system in the sense described above, with its own state space $\s_{AB}$, group of reversible
transformations $\g_{AB}$, and set of allowed effects.

In contrast to quantum theory, the framework of general probabilistic theories allows many different possible composites
for two given systems $A$ and $B$. Every possible composite $AB$ has a set of minimal physical assumptions that it must satisfy:
\begin{itemize}
\item If $\omega_A\in\s_A$ and $\omega_B\in\s_B$ are two local states, then there is a distinguished state $\omega_A\omega_B\in\s_{AB}$
which is interpreted as the result of \emph{preparing $\omega_A$ and $\omega_B$ independently
on the subsystems $A$ and $B$.}
\item If $E_A$ and $E_B$ are local allowed effects on $A$ and $B$, then there is a distinguished allowed effect $E_A E_B$ on $AB$ which is interpreted
as \emph{measuring $E_A$ on $A$ and $E_B$ on $B$ independently,} yielding the total probability that outcome $E_A$ happens on system $A$,
and outcome $E_B$ happens on system $B$.
\item This intuition is mathematically expressed by demanding that
\[
   E_A E_B(\omega_A \omega_B)=E_A(\omega_A) E_B(\omega_B)
\]
where both $E_A E_B$ and $\omega_A \omega_B$ are affine in both arguments.
This also formalizes the physical assumption that the temporal order of the local preparations resp.\ measurements is irrelevant.
\end{itemize}
From the previous point, we can infer that we can represent independent local preparations $\omega_A\omega_B$
and measurement outcomes $E_A E_B$ by tensor products:
\[
   E_A E_B\equiv E_A \otimes E_B,\quad \omega_A\omega_B\equiv \omega_A\otimes\omega_B.
\]
The vector space $V_{AB}$ that carries the composite state space must thus satisfy
\begin{equation}
   \label{eqInclusion}
   V_A\otimes V_B \subseteq V_{AB}.
\end{equation}
For the dimensions of these spaces, we obtain
\begin{equation}
   \label{eqDimensions}
   K_A K_B \leq K_{AB}.
\end{equation}

Now consider two different measurements (for simplicity with two outcomes) $E_B, \bar E_B:=\mathbf{1}_B-E_B$
and $F_B,\bar F_B:=\mathbf{1}_B-F_B$,
where $\mathbf{1}_B$ denotes the trivial effect on system $B$ which yields unit probability on every normalized state. We can think of an agent \emph{Bob},
holding system $B$, who may decide freely (say, according to some local random variable) whether to perform measurement $E_B,\bar E_B$
or $F_B,\bar F_B$.

Suppose that Alice (holding system $A$) performs some measurement after Bob has chosen and performed his measurement on a bipartite state $\omega_{AB}$.
The marginal probability that she obtains (not knowing Bob's outcome) is the same in both cases:
\begin{eqnarray*}
   E_A \otimes \mathbf{1}_B(\omega_{AB})&=& E_A\otimes E_B(\omega_{AB}) + E_A \otimes \bar E_B(\omega_{AB})\\
   &=& E_A\otimes F_B (\omega_{AB}) + E_A\otimes \bar F_B(\omega_{AB}).
\end{eqnarray*}
The same holds with the roles of $A$ and $B$ reversed. This equation follows from our assumptions above on how to represent local measurements.
We have proven that our assumptions imply the \emph{no-signalling property}:
Bob cannot send information to Alice merely by his choice of local measurement (and vice versa). Moreover, the previous
equation shows that the outcome probabilities of all of Alice's measurements are described by the 
\emph{reduced state} $\omega_A:={\rm Id}_A\otimes\mathbf{1}_B(\omega_{AB})$
(note that ${\rm Id}_A$ is the identity transformation, while $\mathbf{1}_B$ is a linear functional). This state corresponds to the marginal
of $\omega_{AB}$ on $A$, and is uniquely characterized by the equation
\[
   E_A(\omega_A)= E_A\otimes\mathbf{1}_B(\omega_{AB})
\]
for all functionals (in particular, all allowed effects) $E_A$.

For physically meaningful composites $AB$, we should demand that reduced states $\omega_A$, $\omega_B$ of all bipartite states $\omega_{AB}\in \s_{AB}$
are valid local states themselves. In fact, we will demand something which is stronger and contains this as a special case. Suppose that
Alice and Bob share $\omega_{AB}$ and Bob performs a measurement and obtains outcome $E_B$. Knowing this outcome leaves
a \emph{conditional} state $\omega^{E_B}_A$ at Alice's side, which by elementary probability theory satisfies
\begin{equation}
   E_A(\omega^{E_B}_A)=\frac{E_A\otimes E_B(\omega_{AB})}{\mathbf{1}_A \otimes E_B(\omega_{AB})}.
   \label{eqConditional}
\end{equation}
We demand that $\omega^{E_B}_A\in\s_A$ for all allowed effects $E_B$ and all $\omega_{AB}\in\s_{AB}$. The reduced
state $\omega_A$ can be written
\[
   \omega_A=\lambda \omega^{E_B}_A + (1-\lambda)\omega^{\bar{E}_B}_A
\]
with $\lambda=\mathbf{1}_A \otimes E_B(\omega_{AB})$; thus, $\omega_A\in\s_A$ by convexity.

In some situations, this condition is automatically satisfied, namely if all effects on $A$ and $B$ are allowed (recall that
not all effects need to be physically possible to implement; above, we have discussed that only a subset of effects
might be physically allowed). The proof will also illustrate that the \emph{cone of unnormalized states} is a useful concept.
\begin{lemma}
Suppose that $A$ and $B$ are state spaces such that all effects are allowed. Then, the inclusion of conditional states in the
local state spaces follows directly from the fact that the composite state space $AB$ contains all product states and effects.
\end{lemma}
\proof
Define the \emph{cone of unnormalized states $A_+$} on $A$ by
\[
   A_+:=\{\lambda\omega_A\,\,|\,\, \omega_A\in\s_A,\lambda\geq 0\}.
\]
Since $\mathbf{1}_A(\lambda\omega)=\lambda$ for $\omega\in\s_A$, a vector $\omega\in A_+$
is a normalized state, i.e.\ $\omega\in\s_A$, if and only if $\mathbf{1}_A(\omega_A)=1$.

The \emph{cone of unnormalized effects} is
\[
   A^+:=\{\lambda E_A\,\,|\,\, E_A(\omega_A)\in[0,1]\mbox{ for all }\omega_A\in\s_A,\lambda\geq 0\}.
\]
Since we have said that all effects are allowed, every linear map $E_A:V_A\to\R$ with $E_A(\omega)\in[0,1]$
is an allowed effect. The set $A^+$ contains all non-negative multiples of those. Both sets $A_+$ and $A^+$ are \emph{closed
convex cones}~\cite{dualconebook}, where ``cones'' refers to the fact that if $x$ is in the set, then $\lambda x$ is also in the set
for all $\lambda\geq 0$.

It is now easy to see that $A^+$ is the ``dual cone'' $(A_+)^*$ of $A_+$, where
\[
   (A_+)^*\equiv \{E:V_A\to\R\,\,|\,\, E(\omega)\geq 0\mbox{ for all }\omega\in A_+\}.
\]
Since $(A_+)^{**} = A_+$, we get also that $A_+$ is the dual cone of $A^+$; in other words,
\[
   A_+=\{\omega\in V_A\,\,|\,\, E(\omega)\geq 0 \mbox{ for all }E\in A^+\}.
\]
Recall the definition of the conditional state in~(\ref{eqConditional}). It follows directly from this
definition that $E_A(\omega^{E_B}_A)\geq 0$ for all allowed effects $E_A$, hence for all $E_A\in A^+$.
But then, we must have $\omega^{E_B}_A\in A_+$. Since $\mathbf{1}_A(\omega^{E_B}_A)=1$, we
get $\omega^{E_B}_A\in\s_A$. The same reasoning holds for $B$ instead of $A$.
\qed

Our state spaces also carry a group of reversible transformations. If $G_A\in\g_A$ is a reversible transformation on $A$,
and $G_B\in\g_B$ one on $B$, it is physically clear that we should be able to accomplish both transformations locally
independently; i.e., $G_A\otimes G_B\in\g_{AB}$. We will assume that composite state spaces satisfy this condition.

One of our postulates below will be the postulate of \emph{local tomography}. This is an additional condition on
composites $AB$ which is sometimes, but not always imposed in the framework of general probabilistic theories: It states that

\begin{center}
\emph{global states are uniquely determined by the statistics of local measurement outcomes.}
\end{center}
Local measurement outcomes correspond to effects of the form $E_A\otimes E_B$.
Thus, the postulate of local tomography states that
$E_A\otimes E_B(\omega_{AB})=E_A\otimes E_B(\varphi_{AB})$ for all $E_A,E_B$ implies that $\omega_{AB}=\varphi_{AB}$.

Since the $E_A$ span the dual space $V_A^*$,
and the $E_B$ span $V_B^*$, the local measurement outcomes span a $(K_A K_B)$-dimensional subspace of $V_{AB}^*$:
\[
   \dim\,{\rm span} \{E_A\otimes E_B\} = (\dim V_A^*)(\dim V_B^*)=K_A K_B.
\]
Any state $\omega_{AB}\in\s_{AB}$ can thus be uniquely specified by $K_A K_B$ linear coordinates
\[
   E_A^{(i)}\otimes E_B^{(j)}(\omega_{AB}),\qquad i=1,\ldots,K_A;\enspace j=1,\ldots,K_B;
\]
in fact, one of these coordinates is redundant, since $\mathbf{1}_A\otimes\mathbf{1}_B(\omega_{AB})=1$,
so $K_A K_B-1$ coordinates are sufficient.
Thus, we obtain an injective affine map from the $(K_{AB}-1)$-dimensional convex set $\s_{AB}$ into $\R^{K_{A} K_B-1}$, which proves that
\[
   K_{AB}-1=\dim\s_{AB}\leq K_A K_B -1.
\]
Due to eq.~(\ref{eqDimensions}), we obtain
\[
   K_{AB}=K_A K_B.
\]
Reading the argumentation backwards shows that this equation is in fact \emph{equivalent} to local tomography, as pointed out by Hardy~\cite{Hardy5}.
It also follows from eq.~(\ref{eqInclusion}) that
\[
   V_{AB}=V_A\otimes V_B.
\]

Thus, we get a certain type of tensor product rule for composite state spaces,
including $\mathbf{1}_{AB}=\mathbf{1}_A\otimes\mathbf{1}_B$.
Note that this is \emph{not} as strong as the tensor product rule of quantum theory, which in addition uniquely specifies the set of global states
on composite systems. In contrast, our tensor product rule only says that the surrounding vector spaces satisfy $V_{AB}=V_A \otimes V_B$,
but does not uniquely specify $\s_{AB}$ in terms of $\s_A$ and $\s_B$. In particular, classical probability theory
satisfies this tensor product rule as well. Suppose that $A$ is a classical bit, and $B$ is a classical $3$-level system. Then the composite $AB$ is classical $6$-level
system, i.e.\ $K_{AB}=6$, while $K_A=2$ and $K_B=3$. We get $K_{AB}=K_A K_B$, which is equivalent to local tomography.

To see that this framework allows for state spaces that are physically very different from quantum theory,
suppose that $A$ and $B$ are both the square state space from Figure~\ref{fig_convexsets}f).
Then, define the global state space $\s_{AB}$ as the set of all vectors $x\in AB$ with $E_A\otimes E_B(x)\in[0,1]$ for all effects $E_A$ and $E_B$,
and $\mathbf{1}_A\otimes \mathbf{1}_B(x)=1$ (normalization). It turns out that this state space contains so-called \emph{PR-box states}
that violate the Bell-CHSH inequality by more than any quantum states~\cite{Barr}. The set of states $\s_{AB}$ itself turns out to be the
eight-dimensional \emph{no-signalling polytope} for two parties with two measurements and two outcomes each. The fact that these state
spaces can have stronger non-locality than quantum theory has been extensively studied~\cite{nav,vandam,boxworld,ic,pr,Barn,Barr} and is
a main reason for the popularity of general probabilistic theories in quantum information.

It is important to keep in mind that the conditions above do not determine the composite state space $\s_{AB}$ uniquely,
even if $\s_A$ and $\s_B$ are given. For example, if $\s_A$ and $\s_B$ are quantum state spaces,
then the usual quantum tensor product is a possible composite $\s_{AB}$, but there are infinitely many other possibilities: one of
them is to define $\s_{AB}$ as the set of unentangled global states. It satisfies all conditions mentioned above.

\subsection{Equivalent state spaces} \label{equivss}
In classical physics, choosing a different inertial coordinate system does not alter the physical predictions of Newtonian mechanics.
A similar statement is true for convex states spaces.

Consider a system $A$, given by a state space $\s_A$, a group of transformations $\g_A$, and some allowed effects.
Suppose that $B$ is another system, and suppose that there is an invertible linear map $L:V_A\to V_B$ such that
\begin{itemize}
\item $\s_B=L(\s_A)$,
\item $E_A$ is an allowed effect on $A$ if and only if $E_A\circ L^{-1}$ is an allowed effect on $B$,
\item $\g_B=L\circ \g_A \circ L^{-1}$.
\end{itemize}
We will then call $A$ and $B$ \emph{equivalent}. Physically, this means that the systems $A$ and $B$ are of the same type in the following
sense. Suppose that we prepare a state $\omega_A$, perform a transformation $T_A$, and finally ask for the occurrence of an effect $E_A$.
The total probability of this is then the same as if we prepare the state $\omega_B=L\omega_A$, perform a transformation $T_B=L\circ T_A\circ L^{-1}$,
and ask for the occurrence of the effect $E_B:=E_A\circ L^{-1}$. In this sense, all physical scenarios on $A$ can be ``translated'' into physical
scenarios on $B$, and vice versa. One may then argue that the linear map $L$ just mediates between two different ways of describing exactly the same type of physical
system. As an example, we may describe the state space of a qubit either as a set of $2\times 2$ density matrices, or as a set of three-dimensional real
vectors, i.e.\ Bloch vectors. These are two different descriptions for exactly the same physics.

Thus, in our endeavor to derive quantum theory, we have to prove that all state spaces satisfying our postulates are equivalent
to quantum state spaces.

\section{The postulates} 
In this section, we describe our postulates and explain their physical meaning. We start with an
axiom on composite state spaces that has already been mentioned in Subsection~\ref{multip_ns} above:

\begin{requirement} [Local tomography] \label{a.correlations} 
	The state of a composite system $AB$ is completely characterized by the statistics of measurements on the subsystems $A,B$.
\end{requirement}

The name ``local tomography'' comes from the interpretation that state tomography on composite systems
can be done by performing local measurements and subsequently comparing the outcomes to uncover correlations.
As already mentioned, this postulate is equivalent to $K_{AB}=K_A K_B$, where $K_A$ denotes the number of degrees of
freedom needed to specify an unnormalized state on $A$.

Our second postulate formalizes a property of physics that physicists intuitively take for granted, and that is in fact
used very often in performing real experiments. Imagine some physical three-level system (that is, with three perfectly distinguishable states and no more: $N=3$)
that we can access in the lab
(it might be quantum, classical, or describable within another theory).
Now suppose that, for some reason, we have a situation where we \emph{never} find the system in the
third of the three distinguishable configurations on performing a measurement.

To have a concrete example, consider a quantum system that consists of three energy levels which can be occupied by
a single particle. Suppose the system is constructed such that the third energy level is actually never occupied (maybe
because the corresponding energy is too high).

The consequence that we expect is the following: \emph{We effectively have
a two-level system.} This is definitely true for quantum theory, and
classical probability theory, but it is not necessarily true for other generalized probabilistic theories. In general, for any number of levels
(perfectly distinguishable states) $N$, we expect to have a corresponding state space $\s_N$. And the collection of states
$\omega\in\s_N$ which has probability zero to be found in the $N$-th level upon measurement should be equivalent to $\s_{N-1}$.

In actual physics, this property is used all the time: We apply ``effective descriptions'' of physical systems, by ignoring impossible
configurations. Qubits manufactured in the lab usually actually correspond to two levels of a system with many more energy levels,
set up in a way such that the additional energy levels have probability close to zero to be occupied.

One may argue that practicing physics would be very difficult if this property did not hold: we would then possibly have to take into account
unobservable potential configurations even if they are never seen. Their presence or absence would affect the resulting state space that we actually observe.
The following ``subspace postulate'', first introduced by Hardy~\cite{Hardy5}, formalizes this idea. It is actually somewhat
stronger than our discussion motivates: it also implies that, for every $N$, there is a \emph{unique} type of $N$-level system $\s_N$.

The notions of complete measurements and equivalent state spaces were defined in Subsections~\ref{SubsecStatesMeasurements} and \ref{equivss}.

\begin{requirement}[Equivalence of subspaces]
\label{a.subspaces}
Let $\s_N$ and $\s_{N-1}$ be systems
with capacities $N$ and $N-1$, respectively.
If $E_1,\ldots, E_N$ is a complete measurement on $\s_N$, then the set of states $\omega\in\s_N$
with $E_N(\omega)=0$ is equivalent to $\s_{N-1}$.
\end{requirement}

The notion of equivalence needs some discussion. Postulate~\ref{a.subspaces} states the equivalence of $\s_{N-1}$ and
\begin{equation}
   \label{S'}
   \s'_{N-1}:=\{\omega\in\s_N\,\,|\,\, E_N(\omega)=0\}.
\end{equation}
Denote the real linear space which
contains $\s_N$ by $V_N$; define $V_{N-1}$ analogously, and set $V'_{N-1}:={\rm span}(\s'_{N-1})$.
Equivalence means first of all that there is an invertible linear map $L:V_{N-1}\to V'_{N-1}$ such that $L(\s_{N-1})=\s'_{N-1}$.
But it also means that transformations and measurements on one of them can be implemented on the other. We now describe
in more detail what this means.

Every effect $E$ on $\s_N$ defines an effect on $\s'_{N-1}$ by restricting it to the corresponding linear space,
resulting in $E\upharpoonright{V'_{N-1}}$. Equivalence implies that the resulting set of effects is in one-to-one
correspondence with the set of effects on $\s_{N-1}$, as described in Subsection~\ref{equivss}.

The transformations on $\s'_{N-1}$ are defined analogously. To be more specific, define $\bar\g'_{N-1}$ as the set
of transformations in $\s_N$ that preserve $\s'_{N-1}$ (or, equivalently, $V'_{N-1}$):
\[
   \bar\g'_{N-1}:=\{T\in\g_N\,\,|\,\, T\s'_{N-1}=\s'_{N-1}\}.
\]
The set of reversible transformations $\g'_{N-1}$ is defined as the restriction of all these transformations to $\s'_{N-1}$
(or rather, as linear maps, to $V'_{N-1}$):
\[
   \g'_{N-1}=\left\{ T\upharpoonright V'_{N-1}\,\,|\,\, T\in \bar\g'_{N-1}\right\}.
\]
Equivalence means that
\[
   \g'_{N-1}=L\circ \g_{N-1}\circ L^{-1}.
\]
Concretely, if $U\in\g_{N-1}$ is any reversible transformation on a state space of capacity $N-1$, then the
transformation $\tilde U:=L\circ U \circ L^{-1}$ is a reversible transformation on $\s'_{N-1}$, i.e.\ $\tilde U \in\g'_{N-1}$.
As such, it can be written $\tilde U=T\upharpoonright \s'_{N-1}$ for some reversible transformation $T\in\g_N$.

It is important to note that we \emph{don't have full information on $T$} -- that is, our postulate does not specify $T$ uniquely,
given $\tilde U$. By definition, $T$ preserves $\s'_{N-1}$ and therefore the subspace $V'_{N-1}$, but
we do not know how it acts on the complement of that subspace -- it might act as the identity there, or it might
have a non-trivial action. Postulate~\ref{a.subspaces} does not specify this. In general, there may (and will) be different $T$
which implement the same $\tilde U$ on the subspace.

Using Postulate~\ref{a.subspaces} iteratively, we see that state spaces of smaller capacity are included (in the sense
described above) in those of larger capacity; symbolically,
\[
   \s_1\subsetneq \s_2 \subsetneq \s_3\subsetneq\ldots
\]

Our next postulate describes the idea that any actual physical theory of probabilities must allow
for ample possibilities of reversible time evolution. In situations where ``no information is lost'' -- assuming
that this situation applies to closed systems --, these systems $A$ must evolve reversibly, that is, according to some
subgroup of the group of reversible transformation $\g_A$. Clearly, if this group is trivial (contains only the
identity), physics becomes ``frozen'': no reversible time evolution is possible at all.

Postulate~\ref{a.symmetry} proclaims a minimal amount of transformational richness for reversible time evolution:
as a minimal requirement, it states that the group of reversible transformations should act transitively on the
pure states. That is, if we prepare a pure state $\omega$, and $\varphi$ is another (desired) pure state on the same
state space, then there should be a reversible transformation $T$ which maps $\omega$ to $\varphi$:

\begin{requirement}[Symmetry]\label{a.symmetry}
   For every pair of pure states $\omega,\varphi\in \s_A$, there is a reversible transformation $T\in\g_A$ such that $T\omega=\varphi$.
\end{requirement}

It is easy to see that Postulate~\ref{a.symmetry} is true for quantum theory: every pure state can be mapped to every other by some
unitary. This example also shows that Postulate~\ref{a.symmetry} is rather weak: in quantum theory, even tuples of perfectly distinguishable
pure states $\omega_1,\ldots,\omega_n$ can be mapped to other tuples $\varphi_1,\ldots,\varphi_n$ by suitable unitaries. This is a much higher
degree of symmetry than what is demanded by Postulate~\ref{a.symmetry}.

There is one postulate remaining. As we discussed in Subsection~\ref{SubsecStatesMeasurements}, given some state space $\s_A$,
not all effects (i.e.\ linear functionals on $A$ which are non-negative on $\s_A$) may be physically allowed. Similarly as for superselection
rules, it might be true that some effects are impossible to implement (an example would be a state space that allows only noisy measurements,
and no outcome whatsoever occurs with probability zero).

In order for our axiomatization to work, we need to exclude this possibility: we postulate that all mathematically
well-defined effects correspond to allowed measurement outcomes. As it turns out, it is sufficient to postulate this for a $2$-level
system $\s_2$ (i.e.\ a generalized bit). In combination with the other postulates, it follows for all other state spaces.

\begin{requirement}[All measurements allowed] \label{a.effects} 
All effects on $\s_2$ are outcome probabilities of possible measurements.
\end{requirement}

From a mathematical point of view, this postulate could also be regarded as a background assumption: structurally, it says that
the class of considered theories is the class of models where the effects are automatically taken as the dual of the states. In other words,
it means that whenever we refer to ``measurements'' in the other postulates, we actually refer to collections of effects without considering
the possibility that additional physical conditions might prevent their implementation.

It is interesting to note that Postulate~\ref{a.effects} can be replaced by a different formulation, which has first been suggested
in the axiomatization by G.\ Chiribella et al.~\cite{GiulioAxioms}. It refers to ``completely mixed states'',
which are states that are in the relative interior of the convex set of states:

\bigskip\noindent
{\bf Postulate 4' (Ref.~\cite{GiulioAxioms}).}  If a state is not completely mixed, then there exists at least one state that can be perfectly distinguished from it.

\section{How quantum theory follows from the postulates}
\label{sqt}
We are now ready to carry out the reconstruction of quantum theory (QT) from the postulates. As it turns out, there will be another solution to Postulates 1.-4., which
is classical probability theory (CPT). By this we mean the theory where the states are finite probability distributions, and the reversible transformations are the
permutations. Figure~\ref{fig_convexsets}a)-c) shows what classical probability distributions look like in terms of convex sets: they are simplices.

Therefore, we will now prove the following theorem:
\begin{theorem}[Main Result]
\label{TheMain}
The only general probabilistic theories, satisfying Postulates 1.-4.\ above, are equivalent to one of the following two theories:
\begin{itemize}
\item \textbf{Classical probability theory (CPT)}: The state space is the set of probability distributions,
\[
   \s_N=\{(p_1,\ldots,p_N)\,\,|\,\, p_i\geq 0,\sum_i p_i=1\},
\]
and the reversible transformations $\g_N$ are the permutations on $\{1,\ldots,N\}$.
\item \textbf{Quantum theory (QT):} The state space $\s_N$ is the set of density matrices on $N$-dimensional complex Hilbert space,
\[
   \s_N=\left\{\rho\in\mathbb{C}^{N\times N}\,\,|\,\, \rho\geq 0,\enspace \Tr\rho=1\right\},
\]
and the group of reversible transformations $\g_N$ is the projective unitary group, that is, the set
of maps $\rho\mapsto U \rho U^\dagger$ with $U^\dagger U =\mathbf{1}$.
\end{itemize}
\end{theorem}
In both cases, all effects must be allowed. Working out the set of effects (that is, linear functionals on states
yielding values between $0$ and $1$), one easily recovers the usual measurements of CPT and QT.

In this paper, we will not give the full reconstruction in all details; the full proof can be found in our more technical paper~\cite{MM}.
Instead, we will try to give a self-contained summary of the reconstruction, its main ideas, and some interesting observations
in the course of the argument.

Before starting to do this, let us discuss a simple observation regarding Theorem~\ref{TheMain}. In order to rule out CPT -- and hence
to single out QT uniquely -- we can tighten Postulate~\ref{a.symmetry} by replacing it with the following modification:\\

\noindent
\textbf{Postulate 3C} (Continuous symmetry.) \emph{For every pair of pure states $\omega,\varphi\in\s_A$, there is a continuous family
of reversible transformations $\{G_t\}_{t\in[0,1]}$ such that $G_0 \omega = \omega$ and $G_1\omega=\varphi$.
}

In other words, every pure state can be ``continuously moved'' into every other pure state. A statement like this is expected to
be true in physical systems with continuous reversible time evolution -- which is the case that seems to be true, to good approximation,
in our universe. The consequence is:

\begin{center}
\emph{The only general probabilistic theory that satisfies Postulates 1, 2, 3C, and 4, is quantum theory (QT).}
\end{center}

\subsection{Why bits are balls}
In QT, the state space of a 2-level system (that is, a generalized bit, or qubit, $\s_2$) is a three-dimensional ball,
the Bloch ball. In CPT, the (classical) bit instead is a line segment, as shown in Figure~\ref{fig_convexsets}. In fact,
this is a ball, too: it is a one-dimensional unit ball. However, quantum $N$-level systems with $N\geq 3$ are not
balls: they contain mixed states in their topological boundary~\cite{Zyczkowski}.

We will now show that all theories satisfying our postulates must have Euclidean ball states spaces as generalized bits.
The dimension of this ball will not be determined yet; this will be done later on.

\begin{figure}
\begin{tikzpicture}[>=stealth]
	\filldraw[fill=white!90!black, draw=black] (0,0)--++(90:1cm)--++(45:1cm)--++(0:1cm)--++(-45:1cm) --++(-90:1cm)--++(-135:1cm)--++(-180:1cm) --cycle;

	\draw (.8,0) node {\LARGE ${\cal S}_2$};

	\draw[thick] (-.9,.1)--+(45:3.1cm) node[near start,sloped,above] {$E =1$};
	\draw[fill=black] (0,1) circle (.04cm);
	\draw (0.25,0.9) node {$\omega_1$};

	\draw[fill=black] (.7071,1.7071) circle (.04cm);
	\draw (.85,1.5) node {$\omega_2$};

	\draw[fill=black] (.35,1.35) circle (.04cm);
	\draw (1.2,.85) node {$\omega_\mathrm{mix}$};
	\draw[->] (.86,.9) --(.38,1.32); 

	\draw[thick] (2.41421,0)--+(75:1.5cm)--+(255:1cm) node[midway,sloped,below] {$E_e =1$};
	\draw[fill=black] (2.41421,0) circle (.04cm);
	\draw (2,0.1) node {$\omega_e$};

	\filldraw[fill=white!90!black, draw=black] (5.2,.5) circle (1.2cm);

	\draw (4.8,0) node {\LARGE ${\cal S}_2$};

	\draw[thick] (5.2,.5)++(-15:1.2cm) --+(75:1.5cm) --+(255:1cm) node[midway,sloped,below] {$E_e =1$};
	\draw[fill=black] (5.2,.5)++(-15:1.2cm) circle (.04cm);
	\draw (4.8,.6)++(-15:1.2cm) node {$\omega_e$};
\end{tikzpicture}

\caption{Like every compact convex set, the bit state space $\s_2$ contains pure states $\omega_e$ that are exposed -- that is,
there is an effect $E_e$ such that $\omega_e$ is the unique state where this effects attains value $1$. Due to Postulate~\ref{a.subspaces},
this proves that $\s_1$ contains a single state only. Now suppose $\s_2$ had lines in its boundary, as in the left picture. Then we would
analogously find another effect $E$ that attains value $1$ on a non-trivial face. Consequently, Postulate~\ref{a.subspaces} would tell us that
$\s_1$ contains infinitely many states -- a contradiction. Thus, $\s_2$ must be strictly convex as in the right picture. Euclidean ballness
follows from group representation theory.
\label{FigBall}}
\end{figure}
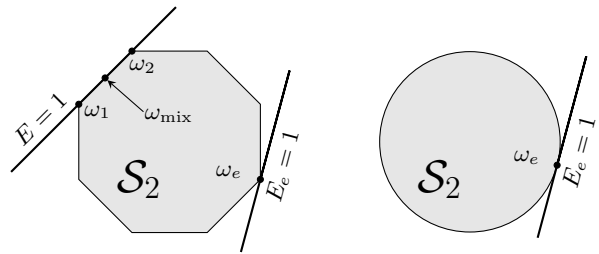

Our argument proceeds in two steps: first, we show that the state space $\s_2$ cannot have lines in its boundary; that is,
we exclude the fact that $\s_2$ has proper faces as in the left picture of Figure~\ref{FigBall}. Using convex geometry language,
we prove that $\s_2$ is \emph{strictly convex}.

As a second step, we show that the symmetry property, Postulate~\ref{a.symmetry}, enforces $\s_2$ to be a Euclidean ball.
The reason for this comes from group representation theory: since the group of transformations acts linearly, there is
an inner product such that all transformations are orthogonal with respect to it.

\begin{lemma}
\label{LemStrictlyConvex}
The state space of the generalized bit $\s_2$ is strictly convex.
\end{lemma}
\proof
Consider any effect $E$ with $0\leq E(\omega)\leq 1$ for all states $\omega\in\s_2$. Then this effect belongs to
a two-outcome measurement (as defined in Subsection~\ref{SubsecStatesMeasurements}), consisting of the
two effects $E$ and $\mathbf{1}-E$. It is important to understand that the level sets $\{x\,\,|\,\, E(x)=c\}$
are hyperplanes of codimension $1$, due to linearity of $E$. This is true for all state spaces $\s$. On the other hand,
given some hyperplane, we can construct a corresponding effect $E$ (with some freedom of offset and scaling) that
has this hyperplane as its level set.

Like every compact convex set, $\s_2$ has at least one pure state $\omega_e$ which is exposed~\cite{Strask} -- that is,
there is a hyperplane which touches the convex set only in $\omega_e$ and in no other point. Thus, we can construct
an effect $E_e$ such that the corresponding hyperplane is $\{x\,\,|\,\, E_e(x)=1\}$, i.e.\ $E_e(\omega_e)=1$, and
$\min_{\omega\in\s_2} E_e(\omega)=0$. But then, $(E_e,\mathbf{1}-E_e)$ distinguishes two states perfectly, which
is the maximal number for a bit -- in other words, this is a \emph{complete measurement}.

Now Postulate~\ref{a.subspaces} says that
\begin{eqnarray*}
   \{\omega\in\s_2\,\,|\,\, (\mathbf{1}-E_e) (\omega)=0\}&=&\{\omega\in\s_2\,\,|\,\,E_e(\omega)=1\}\\
   & =& \{\omega_e\}\simeq \s_1.
\end{eqnarray*}
In other words, $\s_1$ is a trivial state space which contains only a single state. Now suppose that $\s_2$
has lines in its boundary, and therefore non-trivial faces, as depicted on the left-hand side of Figure~\ref{FigBall}.
Then we find a supporting hyperplane that touches $\s_2$ in infinitely many states. Constructing a corresponding
effect $E$ and repeating the argument from above, we analogously argue that $\s_1$ must contain infinitely
many states. This is a contradiction.
\qed

Balls do not have lines in their boundary, but there are many other strictly convex sets -- for example, imagine a
droplet-like figure. However, Postulate~\ref{a.symmetry} says that there is lots of symmetry in the state space $\s_2$:
all pure states (which we now know means all states in the topological boundary due to Lemma~\ref{LemStrictlyConvex}) are connected by reversible transformations.

From this, one can prove that
\begin{lemma}
\label{LemBallness}
The state space $\s_2$ is equivalent to a Euclidean ball (of some dimension $d_2:=K_2-1$).
\end{lemma}

Recall that we denote the dimension of the set of \emph{unnormalized} states by $K_N$; therefore,
the set of normalized states $\s_N$ has dimension $K_N-1$. We will not prove Lemma~\ref{LemBallness} here, but
only sketch where it comes from. An important notion turns out to be the \emph{maximally mixed state}.
On any state space $\s_N$, define $\mu_N$ as a mixture over the group of transformations,
\[
   \mu_N:=\int_{\g_N} G\omega\, dG,
\]
where $\omega\in\s_N$ is any pure state. This is an integral over the invariant measure of the group; see~\cite{group_book,Haar-book} for details of its definition.
It follows from the connectedness of all pure states (Postulate~\ref{a.symmetry}) that
$\mu_N$ does not depend on the choice of the pure state $\omega$. Moreover, $\mu_N$ turns out to be the unique state which
is invariant with respect to all reversible transformations,
\[
   G\mu_N=\mu_N\qquad\mbox{for all }G\in\g_N.
\]
All states $\omega\in\s_N$ span an affine space of dimension $K_N-1$. We can now consider $\mu_N$ to be the origin of that affine space,
turning it into a linear space. Then reversible transformations $G\in\g_N$ act linearly; they preserve the origin. States $\omega$ are represented
by their difference vectors $\hat\omega:=\omega-\mu_N$ that live in this linear space. If a reversible transformation $T$ maps $\omega$ to $\varphi$,
then it also maps $\hat\omega$ to $\hat\varphi$.
By group representation theory, there is an inner product on this linear space
which is invariant with respect to all reversible transformations. As a consequence, if $\omega$ and $\varphi$ are arbitrary pure states,
then there is a reversible transformation $T$ such that $T\hat\omega=\hat\varphi$ due to Postulate~\ref{a.symmetry}, and
so $\|\hat\omega\|=\|\hat\varphi\|$ for the norm corresponding to this inner product.
In the case of a bit, i.e.\ $N=2$, strict convexity implies that we obtain the full Euclidean ball, with the pure states on the surface
and the maximally mixed state $\mu_N$ in the center.

\subsection{The multiplicativity of capacity}
So far, we know that if we combine two state space $A$ and $B$, the joint state space has
dimension $K_{AB}=K_A K_B$ -- this is due to Postulate~\ref{a.correlations}, local tomography, as discussed in Subsection~\ref{multip_ns}.
However, we do not yet know whether the same equality is true for capacity $N$. An important step in the derivation of quantum theory is to prove this.
As it turns out, a key insight is that the maximally mixed state must be multiplicative: if we have two state spaces $A$ and $B$, then the maximally
mixed state on the composite system $AB$ (assuming our postulates) is
\[
   \mu_{AB}=\mu_A\otimes\mu_B.
\]
This is easily proved from the fact that $\mu_{AB}$ must in particular be invariant with respect to all \emph{local} reversible transformations,
leaving $\mu_A\otimes\mu_B$ as the only possibility. A further key lemma is the following:

\begin{lemma}
\label{LemnN}
If there are $n$ perfectly distinguishable pure states $\omega_1,\ldots,\omega_n\in\s_N$ that average to the maximally mixed state, i.e.\
\[
   \mu_N=\frac 1 n \sum_{i=1}^n \omega_i,
\]
then $n=N$.
\end{lemma}
\proof
Clearly, $N\geq n$, since $N$ is the maximal number of perfectly distinguishable states. On the other hand, let $\varphi_1,\ldots,\varphi_N$
be a set of perfectly distinguishable pure states on $\s_N$, and $E_1,\ldots,E_N$ the corresponding effects, i.e.\ $E_i(\varphi_j)=\delta_{ij}$.
Since $1=\sum_{i=1}^N E_i(\mu_N)$, there must be some $k$ such that $E_k(\mu_N)\leq 1/N$. By Postulate~\ref{a.symmetry}, there is
a reversible transformation $G\in\g_N$ with $G\omega_1=\varphi_k$. Thus
\begin{eqnarray*}
   \frac 1 N &\geq& E_k(\mu_N) = E_k\circ G(\mu_N)=\frac 1 n \sum_{i=1}^n E_k\circ G(\omega_i)\\
   &\geq& \frac 1 n E_k\circ G(\omega_1)=\frac 1 n.
\end{eqnarray*}
Thus, we also have $N\leq n$, proving the claim.
\qed

In quantum theory, the maximally mixed state on an $N$-dimensional Hilbert space is the density matrix
\[
   \mu_N=\frac{\mathbf{1}_N} N = \frac  1 N \sum_{i=1}^N |\psi_i\rangle\langle\psi_i|,
\]
if $|\psi_1\rangle,\ldots,|\psi_N\rangle$ denotes an orthonormal basis of $\mathbb{C}^N$ -- that is, if these are pure
states that are perfectly distinguishable. This is in agreement with Lemma~\ref{LemnN}. Moreover, we can prove that
an analogous formula holds for every theory satisfying our Postulates 1.-4.:

\begin{lemma}
\label{LemMaxMixture}
For every $N$, there are $N$ pure perfectly distinguishable states $\omega_1,\ldots,\omega_N\in\s_N$ such that
\[
   \mu_N=\frac 1 N \sum_{i=1}^N \omega_i.
\]
\end{lemma}
We only sketch the proof here: For $N=1$, the statement is trivially true, since $\s_1$ contains only a single state.
For $N=2$, we know that $\s_N$ is a Euclidean ball, with the maximally mixed state in the center. Thus, taking
$\omega_1$ and $\omega_2$ as two antipodal points on the ball (say, north and south pole), we get
\[
   \mu_2=\frac 1 2(\omega_1+\omega_2),
\]
and these states are perfectly distinguishable by an analogue of a quantum spin measurement.
Now consider a generalized bit $A$, and $k$ copies of this physical system denoted $A_1,\ldots,A_k$.
We can form a joint system $A^{(k)}:=A_1 A_2\ldots A_k$; since we do not yet know that we have associativity
of composition, we mean by this $((A_1 A_2) A_3) A_4\ldots$.
Then the maximally mixed state on the resulting state space is
\[
   \mu_{A^{(k)}} = \mu_2\otimes\ldots\otimes\mu_2=\frac 1 {2^k} \sum_{i_1,\ldots,i_k=1,2} \omega_{i_1}\otimes\ldots\otimes\omega_{i_k}.
\]
Since in locally tomographic composites, products of pure states are pure, the $\omega_{i_1}\otimes\ldots\otimes\omega_{i_k}$ are all pure states,
and they are perfectly distinguishable by product measurements. Thus, Lemma~\ref{LemnN}  shows that the capacity of $A^{(k)}$ must
be $N_{A^{(k)}}=2^k$. This proves Lemma~\ref{LemMaxMixture} for all $N$ which are a power of two.
For all other $N$, the lemma is proved by using the fact that $\s_N$ is embedded in some $A^{(k)}$ for some $k$ large enough
due to Postulate~\ref{a.subspaces}, and then constructing the maximally mixed state on $\s_N$ in a clever way from that on $A^{(k)}$.

Now we can form the tensor product of the equations
\[
   \mu_{N_A}=\frac 1 {N_A}\sum_{i=1}^{N_A}\omega_i^A\quad\mbox{ and }\quad\mu_{N_B}=\frac 1 {N_B}\sum_{j=1}^{N_B} \omega_j^B,
\]
and we obtain
\[
   \mu_{N_{AB}} = \mu_{N_A}\otimes\mu_{N_B}=\frac 1 {N_A N_B}\sum_{i=1}^{N_A}\sum_{j=1}^{N_B} \omega_i^A\otimes \omega_j^B,
\]
and Lemma~\ref{LemnN} tells us that capacity must be multiplicative:
\begin{lemma}
$N_{AB}=N_A N_B$.
\end{lemma}
Why is this equation so important? As noticed by Hardy~\cite{Hardy5}, it allows us to draw a surprising conclusion. Every state space $\s_N$ has
unnormalized dimension $K_N$. Since $K_{AB}=K_A K_B$ and $N_{AB}=N_A N_B$ for all state spaces $A$ and $B$ due to our postulates, we get the
following facts:
\begin{itemize}
\item The function $N\mapsto K_N$ maps natural numbers to natural numbers, and is strictly increasing due to Postulate~\ref{a.subspaces}.
\item It satisfies $K_{N_1 N_2}= K_{N_1} K_{N_2}$, and $K_1=1$.
\end{itemize}
As shown in~\cite{Hardy5}, these simple conditions imply that there must be an integer $r\geq 1$ such that
\begin{equation}
   K_N=  N ^r.
   \label{eqr}
\end{equation}
Now recall that the dimension of the bit state space (which is a Euclidean ball) is $d_2:=K_2-1$. It follows that
\[
   d_2 \in \{1,3,7,15,31,\ldots\}
\]
since $K_2=2^r$ for some $r\in\mathbb{N}$. Thus, we see in particular that the bit state space is an \emph{odd}-dimensional Euclidean ball.
The next subsection will deal with the case $d_2=1$; as we will see, this case corresponds to classical probability theory.

\subsection{How to get classical probability theory (CPT)}
Suppose that $d_2=K_2-1=1$; that is, the generalized bit is a one-dimensional ball, as shown in Figure~\ref{fig_convexsets}. A line segment
like this describes a classical bit. What can we say about $N$-level systems for $N\geq 3$ in this case? Equation~(\ref{eqr}) tells us that
the parameter $r$ must be $r=1$, and thus
\[
   K_N=N
\]
for all $N$, not only for $N=2$.

Choose $N$ perfectly distinguishable pure states $\omega_1,\ldots,\omega_N\in\s_N$, and $E_1,\ldots,E_N$ the corresponding
effects with $E_i(\omega_j)=\delta_{ij}$ as well as $\sum_i E_i=\mathbf{1}$. It is easy to see that the states must be linearly independent;
since $K=N$, they span the full unnormalized state space.

Thus, every state $\omega$ can be written $\omega=\sum_{i=1}^N \alpha_i \omega_i$, with $\alpha_i\in\R$ and $\sum_i \alpha_i=\mathbf{1}(\omega)=1$.
But then, $E_j(\omega)=\alpha_j\geq 0$, and so this decomposition of $\omega$ is in fact a convex decomposition.

In other words, the full state space $\s_N$ is a convex combination of $\omega_1,\ldots,\omega_N$ -- that is, a classical simplex
as in Figure~\ref{fig_convexsets}a)--c). These are exactly the state spaces of CPT. Moreover, since for $N=2$, we can permute the
two pure states due to Postulate~\ref{a.symmetry}, we can use the subspace postulate to conclude
that every pair of pure states on $\s_N$ can be interchanged. These transpositions generate the full permutation group, which must thus
be the group of reversible transformations $\g_N$. We have therefore proven the following:

\begin{center}
\emph{In the case $d_2=1$, we get classical probability theory as the unique solution of Postulates 1.-4.}
\end{center}

\subsection{The curious $7$-dimensional case}
Let us now consider the remaining cases, i.e.\ the cases where the dimension of the Euclidean bit ball is
$d_2=K_2-1\in\{3,7,15,31\ldots\}$. The generalized bit carries a group of reversible transformations $\g_2$; by
our background assumptions mentioned in Subsection~\ref{s.trans}, this must be a topologically closed
matrix group. Since it maps the unit ball into itself, it must be a subgroup of the orthogonal group.
Closed subgroups of Lie groups are Lie groups; therefore, $\g_2$ is itself a Lie group.

Denote by $\g_2^0$ the connected component of $\g_2$ containing the identity matrix. We have
\[
   \g_2^0\subseteq SO(d_2).
\]
We know from Postulate~\ref{a.symmetry} that for every pair of pure states $\omega,\varphi\in\s_2$, there is a reversible
transformation $T\in\g_2$ with $T\omega=\varphi$. In other words, $\g_2$ acts \emph{transitively} on the unit sphere, that is,
the surface of the unit ball. It can be shown that this implies that $\g_2^0$ is itself transitive on the unit sphere.

At first sight, it seems that this enforces $\g_2^0$ to be the full special orthogonal group $SO(d_2)$, but this
intuition is wrong. For example, the group of $4\times 4$-matrices
\[
   \left\{ \left.\left(\begin{array}{cc} {\rm re}\, U & {\rm im}\, U \\ -{\rm im}\, U & {\rm re}\,U \end{array}\right)\,\,\right|\,\, U\in SU(2)\right\}
\]
acts transitively on the surface of the $4$-dimensional unit ball, even though it is a proper subgroup of $SO(4)$. The set of all compact connected Lie matrix groups which act transitively
on the unit sphere has been classified in~\cite{MontgomerySamelson,Borel,Onishchik1,Onishchik2}. In general, there are many
possibilities. Fortunately, however, we have additional information: we know that the bit ball has \emph{odd dimension} $d_2:=K_2-1$. It turns out
that there remain only two possibilities:
\begin{itemize}
\item If $d_2\neq 7$, then $\g_2^0=SO(d_2)$.
\item If $d_2=7$, then $\g_2^0$ is either $SO(7)$ or of the form $M G_2 M^{-1}$, where $M$ is a fixed orthogonal matrix, and
$G_2$ is the fundamental representation of the exceptional Lie group $G_2$.
\end{itemize}
In fact, $d_2=7$ appears in our list of possible dimensions of the bit ball, because $7=2^3-1$. In our endeavor to derive quantum theory
from Postulates 1.-4., we will have to show that all the cases $d_2\in\{7,15,31,\ldots\}$ violate at least one postulate. Thus, we see
that the case $d_2=7$ has to be (and is) treated separately.

The appearance of $d_2=7$ as a special case seems like an almost unbelievable coincidence. Is there some deeper significance to this case?
Might there be some interesting unknown theory waiting to be discovered which has $7$-dimensional balls as bits and the exceptional Lie group
$G_2$ as the analogue of local unitaries? We do not know.

\subsection{Subspace structure and $3$-dimensionality}
Having discussed the case of classical probability theory with bit ball dimension $d_2=1$,
the remaining cases are
\[
   d_2\in\{3,7,15,31,\ldots\}.
\]
We will now show that all dimensions $d_2\geq 7$ are incompatible with the postulates, leaving only
the case $d_2=3$ -- that is, the Bloch ball of quantum theory. For the rest of this chapter, we
ignore the special case $d_2=7$ with $\g_2^0=M G_2 M^{-1}$ and $G_2$ the exceptional Lie group; it can be ruled out by an analogous argument.

In the following, we will parametrize the single bit state space as
\[
   \s_2=\left\{\left(\begin{array}{c} 1 \\ \hat\omega\end{array}\right)\,\,|\,\, \hat\omega\in\R^{d_2}, \|\hat\omega\|\leq 1\right\}.
\]
The maximally mixed state becomes $\mu=(1,\mathbf{0})^T$, where $\mathbf{0}\in\R^{d_2}$ denotes the zero vector.
Let $n:=(1,0,\ldots,0)^T\in\R^{d_2}$, then we have two pure states $\omega_1:=(1,n)^T\in\s_2$
and $\omega_2:=(1,-n)^T\in\s_2$, corresponding to the north and south pole of the ball. These states are pure, and they are perfectly distinguished
by the measurement consisting of the two effects (for $\omega\in\s_2$)
\begin{eqnarray*}
   E_1(\omega)&:=& (1+\langle \hat\omega,n\rangle)/2,\\
   E_2(\omega)&:=&(1-\langle\hat\omega,n\rangle)/2.
\end{eqnarray*}
We know that if we combine two bits into a joint state space,
we obtain a state space of capacity four that we call $\s_{2,2}$. It is equivalent to $\s_4$.
Thus, the product states $\omega_i\otimes\omega_j$ with $i,j=1,2$ represent four perfectly distinguishable
states in $\s_{2,2}$, and the corresponding product effects $E_i\otimes E_j$ constitute a complete measurement.
Recall, however, that the joint state space $\s_{2,2}$ is not fully known so far -- all we know is that
the surrounding linear space is the tensor product of the local spaces. At this stage, we do not yet have
 a complete description of the set of all states in $\s_{2,2}$ or $\s_4$.

Using the subspace postulate twice, i.e.\ Postulate~\ref{a.subspaces}, we obtain that the set of states $\omega$ with
$(E_1\otimes E_1+E_2\otimes E_2)(\omega)=1$ is again equivalent to a single bit. This turns out to be
a surprisingly restrictive requirement that we are now going to exploit.  Denote this set of states by $F$ (it is a face
of the state space $\s_{2,2}$), then
\[
   F=\{\omega\in\s_{2,2}\,\,|\,\, (E_1\otimes E_1+E_2\otimes E_2)(\omega)=1\}\simeq \s_2.
\]
In the following, we will label the two bits by indices $A$ and $B$ for convenience. The group $\g_2=SO(d_2)$
contains a subgroup $\g_2^s$ which leaves the axis containing north and south pole invariant, i.e.
\[
   \g_2^s:=\{G\in\g_2\,\,|\,\, G\omega_1=\omega_1\mbox{ and }G\omega_2=\omega_2\}\simeq SO(d_2-1).
\]
If $R\in SO(d_2-1)$, then its action as an element of $\g_2^s$ is
\[
    \left(1,\omega^{(1)},\ldots,\omega^{(d_2)}\right)^T\mapsto \left( 1,\omega^{(1)},R(\omega^{(2)},\ldots,\omega^{(d_2)})\right)^T.
\]
Suppose we apply one transformation of this kind on each part of a bipartite state $\omega$ locally; that is,
a transformation $G_A\otimes G_B$ with $G_A,G_B\in\g_2^s$. Then we have $(E_1\otimes E_1+E_2\otimes E_2)(\omega)=1$
if and only if $(E_1\otimes E_1+E_2\otimes E_2)(G_A\otimes G_B(\omega))=1$. Thus, this transformation
leaves the face $F$ invariant:
\[
   (G_A\otimes G_B) F =F .
\]
We know that the dimension of the linear span of $F$ is $d_2+1$, since it is equivalent to $\s_2$. We will
now explore in more detail how the transformations $G_A\otimes G_B$ act on the face $F$.
In particular, we are interested in the structure of invariant subspaces.

First, consider a single bit. Its unnormalized states are carried by a real vector space $V_A=\R^{d_2+1}$
that we can decompose in the following way:
\[
   V_A=\R\cdot\left(\begin{array}{c} 1 \\ 0\\  \vdots \\ 0 \end{array}\right) \oplus \R\cdot
   \left(\begin{array}{c} 0 \\ 1\\  \vdots \\ 0 \end{array}\right) \oplus A',
\]
where $A'$ denotes the set of all vectors with first two components zero. Since $\mu=(1,0,\ldots,0)^T$
and $G\mu=\mu$, as well as $\omega_1=(1,1,0,\ldots,0)^T$ and $G\omega_1=\omega_1$ for all $G\in\g_2^s$,
these three subspaces are all invariant.

Consequently, the vector space which carries two bits, $V_{AB}\equiv V_A\otimes V_B$, contains the subspace
$A'\otimes B'$ which is invariant with respect to all transformations $G_A\otimes G_B$
for $G_A,G_B\in\g_2^s$. This defines an action of $SO(d_2-1)\times SO(d_2-1)$ on the subspace $A'\otimes B'$.

With a bit of work, one can show that the face $F$ contains at least one state $\omega$ which has non-zero
overlap with $A'\otimes B'$. Denote the projection of that vector onto this subspace by $\omega_{A'\otimes B'}$.
We know that every $(G_A\otimes G_B)(\omega)$ is a valid state in the face $F$, and its component in the
aforementioned subspace is $(G_A\otimes G_B)(\omega_{A'\otimes B'})$.
Now imagine we apply all the local transformations $G_A\otimes G_B$ to the vector $\omega_{A'\otimes B'}$, and we are interested in the orbit -- that is,
in the set of all vectors that we can generate this way.

If $d_2\geq 4$, then the group $SO(d_2-1)$ has a nice property in terms of group representation theory~\cite{group_book}:
it is irreducible. That is, its action on $\mathbb{C}^{d_2-1}$ does not leave any non-trivial subspaces invariant.
This allows us to draw an important conclusion: it implies~\cite{group_book} that the product group $SO(d_2-1)\times SO(d_2-1)$
is also irreducible. But then, the orbit $(G_A\otimes G_B)(\omega_{A'\otimes B'})$ must span the full space $A'\otimes B'$,
which has dimension $(d_2-1)^2$ -- this is a very large orbit.

In fact, it is too large for the subspace postulate: above, we have concluded from Postulate~\ref{a.subspaces} that the span of the face $F$
(which is preserved by those local transformations) must have dimension $d_2+1$, which is less than $(d_2-1)^2$ if $d_2>3$.
Thus, we obtain a contradiction: if the bit ball has dimension
$d_2\in\{7,15,31,\ldots\}$, it is impossible to combine two bits into a joint state space which satisfies all our postulates.

As it turns out, this
is not true if $d_2=3$: the group $SO(d_2-1)=SO(2)$ leaves the span of $(1,i)\t$ invariant; that is, $SO(2)$ is reducible.
Thus, this case is not ruled out by the reasoning above. In group-theoretic terms, this reducibility
is related to the fact that $SO(2)$ is Abelian. In other words, \emph{the fact that rotations commute in $3-1$ dimensions
can be seen as a possible reason of the fact that the Bloch ball is $3$-dimensional.}

\begin{lemma}
The dimension of the bit ball must be $d_2=3$.
\end{lemma}

We have thus uncovered a group-theoretic explanation why the smallest non-trivial quantum systems
have three mutually incompatible, independent components and not more.
Due to Postulate~\ref{a.effects}, we can find all possible measurements on this state space: all
effects (that is, linear functionals) which yield probabilities in the interval $[0,1]$ correspond to
outcome probabilities of possible measurements. It is easy to see that these effects are in one-to-one
correspondence with the quantum measurements (POVMs) on a single qubit.

Furthermore, we know that the group of reversible transformations contains $SO(3)$, the rotations
of the Bloch ball, which correspond to the unitary transformations on a qubit. At this point, however, we
do not yet know whether $\g_2=SO(3)$ or $\g_2=O(3)$.

\subsection{Quantum theory on $N$-level systems for $N\geq 3$}
In the previous section, we have derived quantum theory for single bits. It remains to show
that our postulates also predict quantum theory for all $N$-level systems with $N\geq 3$.
As before, we only sketch the main proof ideas, and refer the reader to~\cite{MM} for proof details.

For a single bit in state $\omega=(1,\hat\omega)^T$, we can obtain the usual representation
as a density matrix by applying a  linear map $L:\R^4\to \mathbb{C}^{2\times 2}_{sa}$, where
the latter symbol denotes the real vector space of self-adjoint complex $2\times 2$-matrices.
This map $L$ is defined by linear extension of
\[
   L(\omega):=(\mathbf{1}+\hat\omega\cdot\vec\sigma)/2,
\]
where $\vec\sigma=(\sigma_x,\sigma_y,\sigma_z)$ denotes the Pauli matrices. The representation
that we obtain (applying $L$ in a suitable way to effects and transformations as well) is equivalent
in the sense of Subsection~\ref{equivss} to the Bloch ball representation.

If we have the state space $\s_{2,2}$ of two bits, we can use the map $L\otimes L$
to represent states $\omega\in\s_{2,2}$ by self-adjoint $4\times 4$-matrices $L\otimes L(\omega)$.
Recall that we have constructed a face $F$ of $\s_{2,2}$ in the previous subsection.
Analyzing $F$ in a bit more detail, one can show that it contains a family of pure states $\omega_u$,
where $u\in[0,\pi)$, which are mapped by $L\otimes L$ onto
\[
   L\otimes L(\omega_u)=|\psi_u\rangle\langle\psi_u|,
\]
where
\[
   |\psi_u\rangle=\cos\frac u 2 |0\rangle\otimes|0\rangle + \sin\frac u 2 |1\rangle\otimes|1\rangle
\]
for some orthonormal basis $\{|0\rangle,|1\rangle\}$.
This is an entangled quantum state with Schmidt coefficients $\cos(u/2)$ and $\sin(u/2)$. Choosing $u$
appropriately, it can attain any value between $0$ and $1$. Thus, by applying local unitaries
(which corresponds to the $SO(3)$-rotations of the local balls), we can generate all pure quantum states.

Denoting $\s'_{2,2}:=L\otimes L(\s_{2,2})$, we have proven the following:
\begin{lemma}
$\s'_{2,2}$ contains all pure $2$-qubit quantum states as pure states.
\end{lemma}

The next step is somewhat tricky: we have to show that there are no further (non-quantum) states in $\s'_{2,2}$.
The idea is to show that \emph{all quantum effects} are allowed effects on $\s'_{2,2}$. Then, if there were additional
non-quantum states in $\s'_{2,2}$, some of these effects would give negative probabilities, which is impossible.

We know that the product effects are allowed on $\s_{2,2}$. Applying the transformation $L\otimes L$, some of the corresponding
effects in $\s'_{2,2}$ are the maps
\[
   \rho\mapsto \Tr\left( P_1\otimes P_2 \rho\right),
\]
where $P_1$ and $P_2$ are one-dimensional projectors. If $T\in\g_{2,2}\simeq \g_4$ is any reversible
transformation on $\s_{2,2}$, denote the corresponding transformation on $\s'_{2,2}$ by $T'\in\g'_{2,2}$.
It maps states $\rho$ to $T'(\rho)$. Suppose we could show the equation
\begin{equation}
   \Tr(P_1\otimes P_2 T'(\rho))=\Tr((T')^{-1}(P_1\otimes P_2)\rho).
   \label{eqWant}
\end{equation}
Then we would be done: due to Postulate~\ref{a.symmetry}, transformations $T'\in\g'_{2,2}$ can map every pure product
state to every other pure state, in particular, to every pure entangled quantum state. This way, $(T')^{-1}$ in the equation
above would generate all entangled quantum effects from the product effect $P_1\otimes P_2$. This is exactly what we want.

Why does eq.~(\ref{eqWant}) hold? Up to a factor $1/4$, the map $L^{\otimes 2}$ is an isometry: for all $x,y\in\R^4\otimes\R^4$, we have
\[
   \Tr\left(L^{\otimes 2}(x)L^{\otimes 2}(y)\right)=\frac 1 4 \langle x,y\rangle.
\]
Thus, translating eq.~(\ref{eqWant}) from $\s'_{2,2}$ back to $\s_{2,2}$, we have to prove that
\[
   \langle E_1\otimes E_2,T\omega\rangle=\langle T^{-1}(E_1\otimes E_2),\omega\rangle.
\]
This is satisfied if $T^T=T^{-1}$ for all $T\in\g_{2,2}$. In fact, we have
\begin{lemma}
\label{LemOrtho}
All reversible transformations $T\in\g_{2,2}$ act as orthogonal matrices on $\R^4\otimes\R^4$.
\end{lemma}
The proof of this lemma is non-trivial and somewhat surprising: it uses Schur's Lemma from group
representation theory, together with the fact that there exist certain kinds of SWAP and CNOT operations
on two bits. These operations are constructed by using Postulate~\ref{a.subspaces}.

Due to Lemma~\ref{LemOrtho}, all the above argumentation becomes solid: eq.~(\ref{eqWant}) is valid, and we get
\begin{lemma}
$\s'_{2,2}$ is the set of $2$-qubit quantum states, and the allowed effects are the quantum effects.
\end{lemma}

So what about the transformations? First of all, we know that that the transformation group of a \emph{single}
bit must be $SO(3)$ -- it cannot be $O(3)$, because local reflections would correspond to partial transpositions
which generate negative eigenvalues on entangled states. Furthermore,
every transformation $T\in\g_{2,2}$ is a linear isometry on the set of self-adjoint $4\times 4$-matrices
that maps the set of density matrices into itself.

According to Wigner's Theorem~\cite{Bargmann,Wigner}, only unitary and anti-unitary maps satisfy this.
However, due to Wigner's normal form, anti-unitary maps generate reflections in some Bloch ball faces
of the state space, which is impossible due to Postulate~\ref{a.subspaces}.

So $\g_{2,2}$ is a subgroup of the unitary group. Due to Postulate~\ref{a.symmetry}, it maps some pure product
state to an entangled state. In other words, $\g_{2,2}$ contains an entangling unitary, and also all local unitaries.
It is a well-known fact from quantum computation~\cite{Bremner} that these transformations generate the full
unitary group.

We have thus shown
\begin{lemma}
The group of reversible transformations $\g'_{2,2}$ on two bits corresponds to
the unitary conjugations, i.e.\ the maps $\rho\mapsto U\rho U^\dagger$ with $U\in SU(4)$.
\end{lemma}

It is now clear that what we did for two bits can also be done for $n$ bits. Since every $\s_N$ is contained
in some $\s_{2^n}$ for $n$ large enough, we can use the subspace postulate to conclude that every state
space $\s_N$ is equivalent to the quantum $N$-level state space.

\section{Conclusions and outlook}
\label{SecConclusions}
We have shown that the Hilbert space formalism of quantum theory can
be reconstructed from four natural, information-theoretic postulates. We hope that this
reconstruction -- together with other recent axiomatizations~\cite{Hardy5,Daki,GiulioAxioms,Hardy11,Hardy12,Zaopo} -- contributes
to a better understanding of quantum theory, and sheds light on some of the mysterious aspects
of its formalism, such as the appearance of complex numbers or unitaries.

One of the main motivations for this work, as mentioned in the introduction, was to find a ``minimal'' set of postulates,
in the sense that removing or weakening any one of the postulates yields new solutions in addition to quantum theory.
Classifying these additional solutions means to analyze ``quantum theory's closest cousins'': these are theories that
are operationally close to quantum theory, but not described by the Hilbert space (or $C^*$-algebra) formalism.
These theories make physical predictions that differ from quantum theory~\cite{Paterek} and that can be tested experimentally~\cite{Ududec}.

Have we achieved the goal of minimality? The postulate which seems to be the strongest is Postulate~\ref{a.subspaces},
which was called ``Subspace Axiom'' by Hardy~\cite{Hardy5}. In fact, in follow-up work~\cite{DigitalApproach,EntDyn}, we show that
Postulate~\ref{a.subspaces} can be significantly weakened: it can be replaced by the requirements
that generalized bits carry exactly one bit of information and not more, and that the state of any system can be reversibly
encoded in a sufficiently large number of generalized bits. As a further benefit, quantum theory with superselection rules
appears as an additional solution. In particular, continuous reversible interaction is sufficient
to single out $d_2=3$ as the dimensionality of the Bloch ball~\cite{EntDyn}. On the other hand, Postulate~\ref{a.correlations}
seems crucial: removing it yields at least quantum theory over the real numbers as an additional solution.

It is currently an open problem whether classical probability theory and quantum theory are the unique theories
satisfying Postulates~\ref{a.correlations}, \ref{a.symmetry} and~\ref{a.effects}. It seems unlikely that Postulate~\ref{a.effects}
can be dropped: adding restrictions to the possible measurements in quantum theory may allow to construct a counterexample.
Furthermore, all current axiomatizations seem to indicate that some assumption on the group of reversible transformations, as in Postulate~\ref{a.symmetry},
is crucial, since this gives the power of group representation theory and the Euclidean structure of the Bloch ball. Interesting progress
has been made recently by Hardy~\cite{Hardy11}, where the corresponding axiom only postulates the existence of suitable \emph{permutations}.

Thus, we have not yet fully achieved the goal of minimality, but we think that our set of postulates is very close to it.
In particular, having as few background assumptions as possible may yield interesting new state spaces that are overlooked if
the full pictorial background framework of quantum circuits is assumed. For example, one might consider the following weaker
version of Postulate~\ref{a.correlations},

\textbf{Postulate 1'.} For every triple (but not necessarily for every pair) of state spaces $A$, $B$ and $C$, there is a tomographically-local
composite $ABC$ which satisfies all other postulates.

It remains an interesting open problem to find a minimal set of axioms, prove its minimality, and systematically characterize
all theories which satisfy some of these axioms, but not all of them. Besides being of interest in its own right,
thorough understanding of alternative routes that nature might have taken may be of crucial importance for experimental
tests of quantum theory, such as tests for higher-order interference~\cite{ThirdOrder}.

\acknowledgments
Research at Perimeter Institute is supported by the Government of Canada through Industry Canada and by the Province of Ontario through the Ministry of Research and Innovation. LM acknowledges support from CatalunyaCaixa.

\end{document}